\documentclass[11pt,a4paper,twocolumn]{article}
\usepackage[utf8]{inputenc}
\usepackage[T1]{fontenc}
\usepackage{amsmath}
\usepackage{amsfonts}
\usepackage{amssymb,textcomp}
\usepackage[pdftex]{graphicx}
\usepackage[usenames,dvipsnames]{xcolor}
\usepackage{multicol}
\usepackage{textcomp,marvosym}
\usepackage{fixltx2e}
\usepackage{cite}

\usepackage{nameref,hyperref}

\newcommand{\matdel}[1]{}
\newcommand{\e}{\mathrm{e}}

\usepackage{geometry}
\usepackage{setspace}

\usepackage{lineno}

\usepackage{fancyhdr}
\fancypagestyle{plain}{
    \fancyhf{} 
    \fancyfoot[]{ } 
    \fancyfoot[L]{ }
    \fancyfoot[C]{\PageNumber}
    \fancyfoot[R]{ }
    \fancyhead[L]{ }
    \fancyhead[R]{ }
}

\title{Non-selective evolution of growing populations}
\author{Karl Wienand}
\author{Matthias Lechner}
\author{Felix Becker}
\author{Heinrich Jung\thanks{hjung@lmu.de}}
\author{Erwin Frey\thanks{frey@lmu.de}}
\date{}

\begin{document}
\twocolumn[\begin{@twocolumnfalse}
\begin{center}
\LARGE Non-selective evolution of growing populations
\medskip

\large{Karl Wienand$^1$, Matthias Lechner$^1$, Felix Becker$^2$, Heinrich Jung$^{2*}$, and Erwin Frey$^{1\dagger}$}
\end{center}

\normalsize
\medskip
\medskip

\noindent 1.\textit{ Arnold-Sommerfeld.Center for Theoretical Physics and Center for NanoScience, Physics Department, Ludwig-Maximilians-Universit\"at M\"unchen, Theresienstrasse 37, D-80333 Munich, Germany.}

\noindent 2.\textit{ Department of Biology 1, Microbiology, Ludwig-Maximilians-Universit\"at M\"unchen, Grosshaderner Strasse 2-4, D-82152 Martinsried, Germany.}

\medskip
\noindent\small{$^\dagger$frey@lmu.de\\$^*$hjung@lmu.de}
\normalsize

\section*{Abstract}
{
Non-selective effects, like genetic drift, are an important factor in modern conceptions of evolution, and have been extensively studied for constant population sizes \cite{kimura_solution_1955,otto_probability_1997}. Here, we consider non-selective evolution in the case of growing populations that are of small size and have varying trait compositions (e.g. after a population bottleneck). We find that, in these conditions, populations never fixate to a trait, but tend to a random limit composition, and that  the distribution of compositions “freezes” to a steady state This final state is crucially influenced by  the initial conditions.  We obtain these findings from a combined theoretical and experimental approach, using multiple mixed subpopulations of two Pseudomonas putida strains in non-selective growth conditions\cite{matthijs_siderophore-mediated_2009} as model system. The experimental results for the population dynamics match the theoretical predictions based on the Pólya urn model \cite{eggenberger_uber_1923} for all analyzed parameter regimes. In summary, we show that exponential growth stops genetic drift. This result contrasts with previous theoretical analyses of non-selective evolution (e.g. genetic drift), which investigated how traits spread and eventually take over populations (fixate) \cite{kimura_solution_1955,otto_probability_1997}. Moreover, our work highlights how deeply growth influences non-selective evolution, and how it plays a key role in maintaining genetic variability. Consequently, it is of particular importance in life-cycles models \cite{melbinger_evolutionary_2010,cremer_evolutionary_2011,cremer_growth_2012} of periodically shrinking and expanding populations.

\bigskip

\bigskip
}
\end{@twocolumnfalse}]

\section*{Introduction}
Stochastic effects play an important role in population dynamics \cite{crow_introduction_1970,hartl_principles_1989,frey_brownian_2005,blythe_stochastic_2007}, particularly when competition between individuals is non-selective. Most previous theoretical analyses have studied how a non-selectively evolving trait can spread and eventually replace all other variants (fixate) under conditions in which the population size remains constant \cite{caballero_developments_1994,otto_probability_1997,gillespie_is_2001}. However, both natural and laboratory populations frequently experience exponential growth. Here we show that genetic diversity in growing populations is maintained despite demographic noise, and reaches a stationary but random limit. We used a well-controlled model system in which well-mixed co-cultures of a wild-type \textit{Pseudomonas putida} strain and an isogenic mutant were grown under non-selective conditions. Multiple subpopulations were generated, each containing a random number of individuals of each strain. Depending on the average initial population size and the strain ratio, we observed distinct stationary probability distributions for their genetic composition. Moreover, we showed that the dynamics of growing populations can be mapped to Pólya urn models \cite{eggenberger_uber_1923}, permitting the observed maintenance of genetic diversity to be understood as the random limit property of a fair game between individual strains. Generalizing the Pólya urn model to include the effects of random initial sampling and exponential growth allowed us to predict the evolution of the composition distribution. Using numerical and analytical methods we found that the distribution broadens at first but quickly “freezes” to a stationary distribution, which agrees with the experimental findings. Our results provide new insights into the role of demographic noise in growing populations.

\section*{Results and discussion}
Evolutionary dynamics is driven by the complex interplay between selective and non-selective (or neutral) effects. The paradigm of non-selective evolution originates from the seminal work of Kimura \cite{kimura_solution_1955}, in which he solved the Wright-Fisher model, thus showing that non-selective effects - and specifically genetic drift - can have a determinant role in evolution. His results sparked an ongoing debate about the nature and potency of randomness as a fundamental evolutionary force \cite{gillespie_is_2001,hahn_toward_2008,hurst_fundamental_2009}. For very small populations genetic drift is generally considered an important factor \cite{gillespie_is_2001}, as the theory successfully predicts the outcomes of neutral evolution experiments \cite{hartl_principles_1989,buri_gene_1956}. 

In most theoretical analyses, constant (or effectively constant) population sizes are assumed, and the role of population growth is neglected. Bacterial populations, however, often undergo rapid growth - especially when they are small. For example, as few as 10 individuals of some highly virulent pathogens (e.g. enterohemorrhagic \textit{Escherichia coli} or \textit{Shigella dysenteriae}) suffice to initiate a deadly infection in a human host \cite{center_for_food_safety_and_applied_nutrition_causes_2012,kurjak_textbook_2006}. Another case of small, growing populations are water-borne bacteria that feed on phytoplankton products. Due to nutrient limitation in open water, these bacteria typically live in small populations in close proximity to the planktonic organism \cite{grossart_marine_2005}. During spring blooms, the phytoplankton releases more organic material, boosting the bacterial growth rate \cite{grossart_marine_2005,bird_empirical_1984,buchan_master_2014}. In nature, such small populations often form by adventitious dispersal from a larger reservoir population \cite{stoodley_biofilms_2002}. A typical example is the spreading of pathogens from host to host. This random “sampling” from a reservoir yields small populations whose genetic compositions differ from that of the reservoir (a phenomenon known as the \textit{founder effect} \cite{levin_dispersion_1974}). Recent studies also showed that the combination of population growth and stochastic fluctuations can have a major impact on the evolution \cite{melbinger_evolutionary_2010,cremer_evolutionary_2011,cremer_growth_2012,sanchez_feedback_2013} and genetics \cite{parsons_consequences_2010} of small populations. 

\begin{figure}[!htbp]
\center\includegraphics{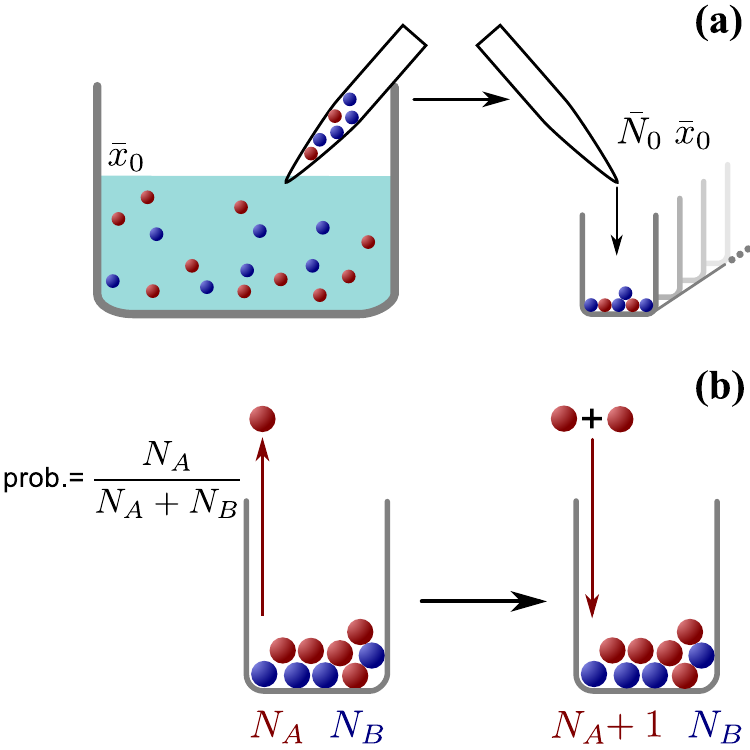}
\caption{\small
Schematic depiction of urn sampling and growth. \textbf{(a)} Schematic illustration of the random initial conditions. An infinite reservoir contains a diluted mixture of bacteria, a fraction $\bar x_0$ of which are of strain $A$. We draw small volumes of liquid from the reservoir containing small, random numbers of individuals, which conform to a Poisson distribution with mean  (determined by the dilution of the reservoir population). A certain fraction of this initial population is of strain $A$. The mean value of this fraction is equal to $\bar x_0$. We use these individuals to initiate populations in the wells of a microtiter plate, so that each population starts with a random size $N_0$ and a random fraction of A-individuals $x_0$. \textbf{(b)} Illustration of the P\'olya urn model. If a bacterial population is represented as an urn, each individual as a marble and each bacterial strain as a color, this urn model captures the essentials of bacterial reproduction in our populations. At each iteration, a marble is drawn at random and returned to the urn, together with another one of the same color. The probability of extracting a marble of either color is determined solely by its relative abundance, making the process non-selective (since no strain has inherent advantages). The rate of growth in population size can be rendered exponential by letting the waiting time between successive iterations be exponentially distributed (also known as \textit{Poissonization}).
\label{fig:setup}}
\end{figure}

To probe how population growth shapes genetic diversity, we used a well-characterized microbial model system, namely the soil bacterium \textit{Pseudomonas putida} KT2440 \cite{buckling_siderophore-mediated_2007,matthijs_siderophore-mediated_2009,cornelis_iron_2009}. The wild-type strain KT2440 produces pyoverdine, an iron-scavenging molecule that supports growth when iron becomes scarce in the environment.
Here we consider co-cultures of two genetically distinct strains: the wild-type, pyoverdine-producing strain KT2440 (strain $A$) and the mutant non-producer strain 3E2 (strain $B$). We set up conditions of non-selective competition between these strains by using an iron-replete medium (casamino acids, supplemented with 200 $\mu$M FeCl$_3$). In this medium, production of pyoverdine is effectively repressed \cite{cornelis_iron_2009}, such that both strains have the same growth rate and neither has an advantage (see SI). Producer (KT2440 wild type) and non-producer (3E2) strains were first mixed and diluted to yield Poisson dilution conditions. Then we initiated a large number of subpopulations from this reservoir by pipetting aliquots of the cell suspension into the wells of a 96-well plate, thereby generating a large ensemble of subpopulations with a random distribution of initial cell number $N_0$ and producer fraction $x_0$  (Figure~\ref{fig:setup}). Use of shaken liquid cultures ensured homogeneous well-mixed conditions for all cells in the same well (access to nutrients, oxygen, etc.), and exponential growth was observed in all cases.

This experimental setting is well described within the mathematical framework of a \textit{Pólya urn model}. Consider each bacterium in the population as a marble in an urn, and its genotype as the color of the marble (e.g. red for strain $A$, and blue for strain $B$). Population growth results from single reproduction events in which an individual randomly divides.  This is mathematically equivalent to a stochastic event in which a marble is chosen at random from the urn and put back, together with another one of the same color. This random process, introduced by Eggenberger and Pólya \cite{eggenberger_uber_1923}, exhibits several important properties \cite{cohen_irreproducible_1976,schreiber_urn_2001,sornette_why_2009,pemantle_survey_2006}.
It is \textit{self-reinforcing}: each time a marble is extracted, another one of the same color is added, increasing the likelihood of extracting a marble of that color again. In the context of bacterial populations, this means that every birth event for one strain makes it more likely that further birth events of that same strain will occur in the future. Note, however, that \textit{fixation}, i.e., complete loss of one type of marble from the population, cannot occur, simply because in the Pólya urn model marbles are neither removed nor do they change their color. This fully reflects the experimental conditions: During exponential growth, rates of cell death are negligible, and within the observation time mutations will be extremely rare, given the population sizes considered. The bacteria in each well reproduce randomly at a per-capita (average) rate $\mu$. To translate this to the urn model, drawing of a marble is assumed to be a stochastic Poisson process, with a “per-marble” rate $\mu$ (a procedure known as \textit{Poissonization} or \textit{embedding} \cite{athreya_embedding_1968,kotz_generalized_2000}).
 Mathematically, the growth process in then described by a Master Equation: The time evolution for the probability $P(N_A,t)$ of finding $N_A$ individuals of strain $A$ at time $t$ reads

\begin{equation}
\frac{\mathrm{d}}{\mathrm{d}t}P(N_A;t) = (N_A-1) P(N_A-1;t) - N_A P(N_A;t)\,,
\label{eq:master-eq}
\end{equation}
where we have set the growth rate to $\mu=1$ in order to fix the time scale (for an introduction to the mathematical concepts see, e.g., \cite{allen_introduction_2003}); the corresponding Master equation for individuals of strain $B$ is of identical form. To study the composition of the populations, we use the more convenient quantities $N=N_A + N_B$ (total size) and $x=N_A/N$ (fraction of individuals of strain $A$).

To start the experiment, we inoculated the wells of 96-well-plates by drawing small volumes of diluted liquid bacterial culture from a large reservoir (Fig.~\ref{fig:setup}\textbf{(a)}). Each volume contains a random number of bacteria whose mean value  is controlled by the dilution of the reservoir. The fraction of bacteria of strain $A$ (wild type) in that volume is also random, with its mean value $\bar x_0$ given by the fraction of strain $A$ in the reservoir. In the mathematical formulation, this setup corresponds to stochastic initial conditions for the Pólya urn model: the initial population size $N_0$ for each well is given by a Poisson distribution with mean $\bar N_0$, and each individual is assigned to strain $A$ or $B$ with probability $\bar x_0$ and $1 - \bar x_0$), respectively. This procedure is also equivalent to treating the initial numbers of $A$- and $B$-individuals as independent, Poisson-distributed random variables with mean values $\bar N_0 \bar x_0$ and $\bar N_0 (1 - \bar x_0)$), respectively \cite{cremer_evolutionary_2011}.

\begin{figure}[!htbp]
\center\includegraphics{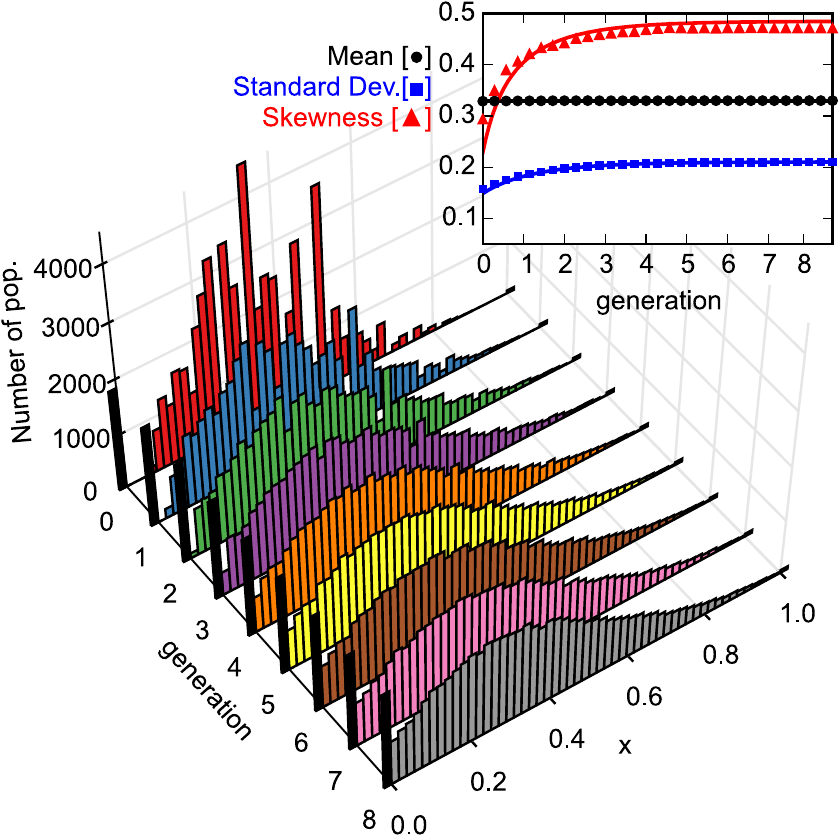}
\caption{\small
Time series for the simulated distribution of the population composition $x$. The distribution initially broadens, then freezes to a steady state (see supporting video). The fraction of populations that have $x=0$ or $x=1$ (indicated by the black bins) remains constant during the time evolution, as expected for a P\'olya urn process, and in contrast to expectations from genetic drift. In each well the population follows a stochastic path and reaches a (random) limit composition, and the distribution freezes only when all populations reached their limit. The parameter values used in the simulation are $\bar N_0 = 10$ and $\bar x_0 = 0.33$ The inset shows the mean, standard deviation and skewness as a function of the number of generations, with symbols denoting numerical simulations, and the solid lines corresponding to the theoretical prediction of Eq~\eqref{eq:variance-evolution} and the analogous ones for the other moments (see SI). Analytical and numerical values agree. The mean $\langle x\rangle$ remains constant throughout the evolution, as expected for a non-selective process; standard deviation and skewness saturate to limit values, confirming the freezing of the distribution.
\label{fig:freezing}}
\end{figure}

Figure~\ref{fig:freezing} shows a time series of the histogram for the composition $x$ of all subpopulations considered, as obtained from a stochastic simulation of the Master Equation~\eqref{eq:master-eq} for a given random initial condition (with $\bar N_0 = 10$ and $\bar x_0 = 0.33$). Surprisingly, the distribution first broadens, but then quickly “freezes” to a steady state (see supporting video).
This is genuinely different from Kimura’s result for populations with constant size \cite{kimura_solution_1955} (or similar results with effectively constant size \cite{otto_probability_1997}) where the balance between stochastic birth and death events leads to genetic drift, and thereby eventually to the extinction of one of the two strains.
In contrast, for a growing population, death events are negligible, and therefore there is no fixation of the population during growth. Instead, fixation arises as a direct consequence of the initial sampling process, as can be seen from the heights of the black bins in the histogram (at $x=0$ and $x=1$), which remain constant over time (Fig.~\ref{fig:freezing}). During growth, the composition of each subpopulation, instead of drifting to fixation at either $x=0$ or $x=1$, reaches a stationary limit value $x^*$, where it remains thereafter \cite{brian_arthur_path-dependent_1987}. This limit value is random: starting several subpopulations (urns) from the exact same initial composition of strain $A$ and $B$ (blue and red marbles), each reaches a limit, but in general these limits differ from one another. Once all of the subpopulations in an ensemble reach their limit, the distribution of the population composition freezes to a steady state, which is equal to the probability distribution of $x^*$. Similar random limit properties appear in other fields, with \textit{lock-in} in economics as the best-known example \cite{sornette_why_2009}. 

The inset in Fig.~\ref{fig:freezing} shows approximate solutions for the time evolution of mean, standard deviation, and skewness of the composition $x$, which we obtained by analytically solving the Master Equation~\eqref{eq:master-eq} (see SI). The analytical results agree well with their numerical counterparts. In particular, the mean value remains constant over time, as it must for a non-selective process. For the time evolution of the variance, which is a measure for the spread of a distribution, we obtain to leading order in population size

\begin{equation}
\text{Var}_\text{poi}[x](t)=\frac{2 - \e^{-t}}{\bar N_0} \bar x_0 (1-\bar x_0) \,.
\label{eq:variance-evolution}
\end{equation}
The broadening and freezing of the distribution is reflected in the exponential decay term of the variance. Note that the skewness increases as well, because growth is self-reinforcing (see inset in Fig.~\ref{fig:freezing}). To further test the validity of the stochastic simulations, we also calculated the limit values of the average and variance after extended periods of evolution exactly, and found that they match the numerical solutions of the Master Equation perfectly (see SI).

We tested these theoretical predictions using \textit{P. putida} as a bacterial model system. We mixed the wild-type and mutant strains in order to obtain different initial fractions $\bar x_0$. The degree of dilution of the mixture determines the average initial cell number $\bar N_0$, with which we inoculated 120 wells per experiment (96-well plate format). In order to compare the experimental data with our model, we set up simulations that matched the experimental configuration by initializing $\bar N_0$ and $\bar x_0$ with the same values as measured in the experiments. We simulated the time evolution of about $10^4$ populations, grouped in “virtual plates” of 120 wells each. Every virtual plate produced a histogram like the one we obtained from experiments. We then generated an average histogram of the virtual plates and used its values to compute the binomial confidence intervals \cite{wilson_probable_1927} for the count in each bin, and compared those with the experimental distribution. 

\begin{figure}[!htbp]
\center\includegraphics{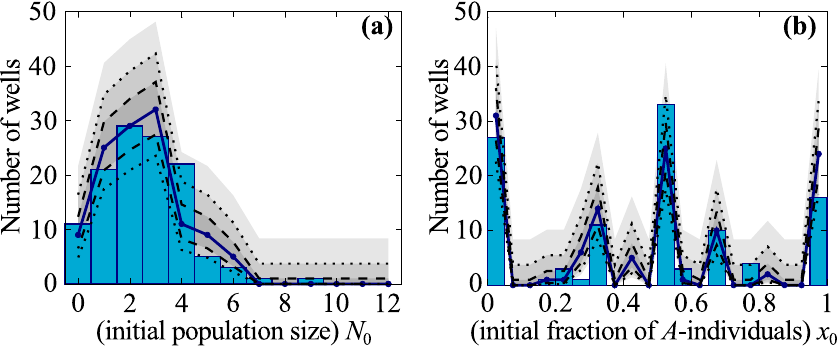}
\caption{\small
Initial distributions for population size $N_0$ and composition $x_0$ (parameter values $\bar N_0=2.55$, $\bar x_0 = 0.45$). The experimental distributions (bars) for $N_0$ (panel \textbf{(a)}) and $x_0$ (panel \textbf{(a)}) are measured from 120-well ensembles. The average $\bar N_0$ and $\bar x_0$ calculated from the measured values determine the parameters for the simulated distributions. The theoretical average distribution (solid blue line) is the average of the same distributions generated for 84 sets of 120 wells. Using that average we calculate the Wilson binomial confidence intervals (gray areas) for 68\% (between dashed lines), 95\% (between dotted lines) and 99.73\% confidence. The measured and simulated distributions agree well within statistical error, confirming our assumption that individuals of strain $A$ and $B$ in the experiments start Poisson-distributed with mean $\bar N_0 \bar x_0$ and $\bar N_0 (1-\bar x_0)$, respectively. The ragged distribution of $x_0$ derives from a small-number effect, and disappears at larger values of $N_0$.
\label{fig:initial-distrib}}
\end{figure}

Figure~\ref{fig:initial-distrib}\textbf{(a)} shows a representative experimental histogram of the initial population sizes $N_0$ for strong dilution with $\bar N_0=2.55$. It is well approximated by a Poisson distribution, and agrees with the simulation results within statistical errors (blue line and shaded gray areas in Fig.~\ref{fig:initial-distrib}\textbf{(a)}). Figure~\ref{fig:initial-distrib}\textbf{(b)} shows the probability distribution of the corresponding initial compositions $x_0$ of the populations, where again theoretical and experimental values agree well within statistical error. Note also that in every well the composition $x_0$ must be a simple fraction; this means that only a few numerical values are possible for small initial population sizes $N_0$. This small-number effect explains why the distribution of $x_0$ in Fig.~\ref{fig:initial-distrib}\textbf{(b)} is so ragged. The distribution becomes much smoother for larger initial population sizes (see SI). Taken together, these results for the distribution of initial population size and composition confirm that the inoculation of the individual wells is a stochastic sampling process with Poissonian statistics.

\begin{figure}[!htbp]
\center\includegraphics{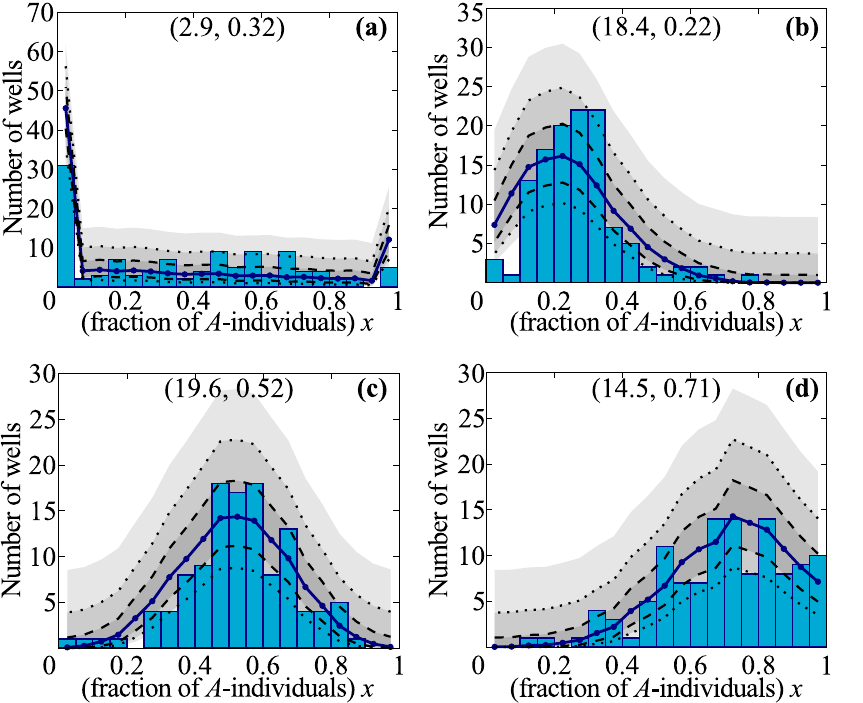}
\caption{\small
Steady-state distributions of population composition $x$ for different initial conditions. The experimental distribution (bars) is the result of growth on 120 independent wells. We use the measured average $x_0$ and $N_0$ from the experiments to initialize the simulations of several 120-well ensembles. After growth, we compute the histogram for each of these ensembles and obtain the average theoretical distribution (blue line). Using the values from this distribution, we compute the three confidence intervals (shaded gray areas) for each bin for 68\% (between dashed lines), 95\% (between dotted lines) and 99.73\% confidence. The two sets of data match: most experimental data agree with the first prediction confidence region, practically all with the second one. The limit distributions are clearly different from the initial ones (see SI). The importance of growth in changing the distributions depends on the initial size $N_0$. Parameter values: $\bar N_0 = 2.9$, $\bar x_0 = 0.32$ (panel \textbf{(a)}); $\bar N_0 = 18.4$, $\bar x_0 = 0.22$ (panel \textbf{(b)}); $\bar N_0 = 19.6$, $\bar x_0 = 0.52$ (panel \textbf{(c)}); $\bar N_0 = 14.5$, $\bar x_0 = 0.71$ (panel \textbf{(d)}).
\label{fig:final-distrib}}
\end{figure}

Next, we were interested in how the composition of the bacterial population would evolve under non-selective (neutral) growth conditions. To this end we let the 120 populations grow for an 11-hour period, during which they remained in exponential growth phase. Then we measured the population size $N(t)$ in each well by counting colony-forming units, and $x(t)$ by counting the pyoverdine-producing colonies (see \nameref{materials}). Figure~\ref{fig:final-distrib} shows the final outcome for four different initial conditions, i.e. combinations of the initial average population size $\bar N_0$ and composition $\bar x_0$. We first wanted to know what determines the number of wells that contain only individuals of either strain $A$ or strain $B$, i.e. that are fixated. To this end we compared the experimentally observed values with the corresponding predictions from the numerical simulations of the Pólya urn model (Fig.~\ref{fig:final-distrib}). Since both results agree within statistical error, we conclude that fixation of a population is a consequence of the initial sampling process and is not due to fixation during population growth. This is especially obvious for small average initial population size or compositions close to $x=0$ or $x=1$, where a large fraction of the wells contains cells of strain $A$ or $B$ only (Fig.~\ref{fig:final-distrib}\textbf{(a)} and Fig.~\ref{fig:final-distrib}\textbf{(d)}). Next we wished to learn how the final distribution of the population composition (i.e. the random limits, $x^*$) depends on the initial average composition $\bar x_0$. For $\bar x_0 = 0.5$, we observed both by experiment and theoretically that the initial distribution significantly broadened (by a factor of $\sqrt{2}$) but remained symmetrical (Fig.~\ref{fig:final-distrib}\textbf{(c)}). In contrast, starting from distributions with average values below or above 0.5 caused the final distribution to broaden and also become skewed towards smaller or larger values of $x$, respectively (Fig.~\ref{fig:final-distrib}\textbf{(b)} and Fig.~\ref{fig:final-distrib}\textbf{(d)}). Moreover, we found quantitative agreement between experiment and numerical simulations within statistical errors in all analyzed parameter regimes (see blue lines and shaded areas in Fig.~\ref{fig:final-distrib}): most experimental histograms fall within the first confidence interval of the prediction (darkest gray areas, between dashed lines), and almost all fall within the 99.73\% confidence interval. 

Taken together, our combined theoretical and experimental analysis gives a coherent picture of evolution during non-selective (exponential) growth.  We have shown, experimentally and by analogy with the Pólya urn model, that for each well-mixed population the composition of the population reaches a random stationary limit , and, unlike populations with constant size, generally does not fixate. For a large ensemble of populations, this implies that the probability distribution for the population composition converges to limit distributions (Fig.~\ref{fig:freezing} and Fig.~\ref{fig:final-distrib}), which are nothing like Kimura’s result for constant-sized populations. Our result is also quite different from that obtained in range expansion experiments \cite{hallatschek_genetic_2007} or other settings featuring population growth without death on two-dimensional substrates. There, monoclonal sectoring patterns arise as a consequence of random genetic drift, which drives population differentiation along the expanding fronts of bacterial colonies, unlike our well-mixed populations that freeze to coexistence.

Our study also shows that, in a growing population with stochastic initial conditions, demographic noise has two possible sources: the initial sampling process by which subpopulations are formed, and the subsequent growth process. The initial average population size $\bar N_0$ sets their relative weight (see SI). For very small $\bar N_0$, of the order of one or two individuals, the formation process already determines the final composition distribution: most populations start off fixated, many with just a single founder individual, and the composition of each well remains the same during growth. For very large $\bar N_0$, of the order of a few hundreds, the sampling process is again central: the composition distribution changes very little before freezing, and growth generates only a very limited amount of variation. In these two limiting regimes, neglecting stochastic effects during growth leaves the evolutionary outcome practically unchanged. In contrast, for small founder colonies such as those typically found during population bottlenecks \cite{kurjak_textbook_2006,grossart_marine_2005,hammerschmidt_life_2014} ($\bar N_0 \sim 10$), population growth is responsible for the major part of the variation observed in the final distribution.

Moreover, our results reveal that a growing population reaches a random limit composition much faster than genetic drift leads to fixation in populations of constant size. Typical fixation times for genetic drift increase logarithmically with the population size \cite{blythe_stochastic_2007}, while the time scale for freezing is independent of population size. This has important consequences for the role of stochastic effects when a population passes from exponential growth phase to stationary phase, in which growth rate and death rate are equal. Then, the composition of the population shows both freezing and fixation, albeit at quite distinct times because the relevant time scales differ markedly. During growth the composition distribution quickly freezes, as described above. Once the population reaches its stationary size, it slowly drifts to fixation, following Kimura-like dynamics.

We also believe that our results have a broad range of applications since growing populations are ubiquitous in nature. For example, experimental studies of \textit{P. aeruginosa} \cite{stoodley_biofilms_2002,hall-stoodley_bacterial_2004} have shown that typical life cycles pass through different steps with regularly occurring dispersal events being followed by the formation of new colonies. As initial colony sizes are typically small, such dispersal events coincide with \textit{population bottlenecks} and subsequent exponential growth. During these phases of the life cycle, population dynamics is often selectively neutral and hence falls within the framework of the present work. The degree of diversity generated during these population bottlenecks has been shown to be crucial for some proposed mechanisms for the evolution of cooperation under selective pressure \cite{melbinger_evolutionary_2010,cremer_evolutionary_2011,cremer_growth_2012,chuang_simpsons_2009,oliveira_evolutionary_2014,melbinger_emergence_2015}. Our analysis quantifies the ensuing degree of diversity and points to the relative importance of sampling versus growth for long-term behavior of the reservoirs. This may have important consequences for the degree of genetic diversity observed in natural populations with life-cycle structures \cite{hammerschmidt_life_2014}.

\section*{Materials and Methods}
\label{materials}
\subsection*{Strains and cultivation conditions}
\small
The \textit{P. putida} strains KT2440 (wild type) and 3E2 (mutant with defective pyoverdine synthesis) \cite{matthijs_siderophore-mediated_2009} were used as pyoverdine producers and non-producers, respectively.
Cells were grown in casamino acid medium (CAA) containing per liter: 5 g casamino acids, 0.8445 g K\textsubscript{2}HPO\textsubscript{4}, 0.1404g MgSO\textsubscript{4}•(H\textsubscript{2}O) \cite{matthijs_siderophore-mediated_2009}.
The CAA medium was supplemented with 200 $\mu$M FeCl\textsubscript{3} (CAA-Fe) to suppress pyoverdine production (see Table S9).
Overnight cultures of the individual strains in CAA-Fe medium were adjusted to an OD\textsubscript{600} of 1, diluted 10\textsuperscript{-2} fold, mixed to yield the desired producer fraction, and further diluted to create Poisson distribution conditions. Producer/non-producer co-cultures were started by inoculating the central 60 wells of two 96-well plates thereby adjusting the average initial cell number $\bar N_0$ to values between 2 and 25 cells/150 $\mu$L. Wells at the border of the plates were filled with water to minimize evaporation from central wells. For non-selective growth, co-cultures were grown in CAA-Fe medium shaking at 30\textdegree C  for given periods of time. Due to the random distribution of initial cell number $N_0$ and producer faction $x_0$ in the 120 wells, each experiment was unique. An experiment was limited to 120 wells to allow initiation of the analysis of the subpopulations in the individual wells without uncontrolled changes of growth parameters during analysis.
The experiment duration was set to 11h to allow evolution to act for a significant number of generations, while leaving bacteria in exponential growth phase (see SI)

\subsection*{Determination of growth parameters}

Cell numbers $N_0$ and $N(t)$ were determined by counting the colony forming units (cfu) of individual wells.
For this purpose 100$\mu$L aliquots of the individual wells were plated on cetrimide \cite{brown_use_1965} or King's B agar (contains per liter: 20 g peptone, 10 g glycerol, 1.965 g K\textsubscript{2}HPO\textsubscript{4}(3H\textsubscript{2}O), 0.842 g MgSO\textsubscript{4}(H\textsubscript{2}O) \cite{king_two_1954}.
Producer fractions $x_0$ and $x(t)$ were determined based on the capability of cells to produce the green fluorescent pyoverdine either by direct counting of fluorescent and non-fluorescent colonies on the plates or after growth in iron-limited CAA medium.
The fraction of dead cells was determined by life/dead staining with propidium iodide \cite{suzuki_dna_1997}, and was always \textless0.02 of the total cell number under the experimental conditions.

\subsection*{Simulation of growing populations}

We performed simulations of 10080 wells using a Gillespie algorithm \cite{gillespie_general_1976}. The initial numbers of “cells” per well were drawn at random from a Poisson distribution with a mean value of $\bar N_0$ measured in the corresponding experiment. The strain assigned to every individual in each well was determined by the outcome of a Bernoulli trial (i.e., coin-flip-like process) and the probability of assignment to strain $A$ was set to the value of $\bar x_0$ measured in the experiment. After initialization, wells were grouped into 84 virtual 120-well “plates”, and a random waiting time was selected for each well, drawn from an exponential distribution with the population size as parameter. The Gillespie algorithm was run until the average size across all wells matched the average size measured at the end of the growth experiments, or until a specified time had elapsed.

\section*{Acknowledgments}
For fruitful discussions we thank Madeleine Opitz, Johannes Knebel, Markus Weber, Stefano Duca, Matthias Bauer, Kirsten Jung. We thank Michelle Eder (HJ lab) for excellent technical assistance. P. putida strain 3E2 was kindly provided by P. Cornelis (Vrije Universiteit Brussels, Belgium).

%
%
%

\newpage
\normalsize
\onecolumn
\begin{center}
\LARGE Non-selective evolution of growing populations\\Supplementary Information
\medskip
\setcounter{page}{1}

\large{Karl Wienand, Matthias Lechner, Felix Becker, Heinrich Jung, and Erwin Frey}
\end{center}

\section*{Video: Time evolution of composition distribution.}
\label{vid:freezing}
\url{https://vimeo.com/108884249}

The distribution of compositions $x$ first broadens due to demographic noise, but soon ``freezes'' to a steady state. The steady state form is maintained as long as the populations grow. Parameter values are $\bar N_0 = 10$, $\bar x_0 = 0.33$ (as for Fig. \ref{fig:freezing}).

\section*{Exact calculations for steady-state composition distribution and moments.}
\label{sup:exact-calculations}
\subsection*{Calculation of probability distribution.}

Each population in the ensemble is initialized with $A_0$ individuals of type $A$ and $B_0$ of type $B$.
In the general case, $A_0$ and $B_0$ are independent random variables for each population with distributions $P(A_0)$ and $P(B_0)$.
All populations evolve for $\Delta N$ reproduction events, of which a random amount $\Delta A$ generate new $A$-individuals.
From the mathematical literature \cite{pemantle_survey_2006}, it is well-known that $\Delta A$ follows a beta-binomial, with $A_0$, $B_0$ and $\Delta N$ as parameters.
The fraction of $A$-individuals $x$, then follows the probability
\begin{equation}
P(x) = \sum_{A_0, B_0} P(A_0) P(B_0) P(\Delta A = x (A_0 + B_0 + \Delta N) - A_0| A_0, B_0, \Delta N)\,,
\end{equation}
where the sums run over all allowed values of their respective indices.
$P(\Delta A = k | A_0, B_0, \Delta N)$ is the probability of $\Delta A$ being equal to $k$, given the values of $A_0$, $B_0$ and $\Delta N$.
The sum may easily be performed numerically.
For the moments of the distribution there are, however, also closed-form analytic expressions.

\subsection*{Exact calculation of asymptotic moment values}

Let $\langle\cdot\rangle_0$ be the average over the initial conditions, $\langle\cdot\rangle_{\Delta A}$ be an average over $\Delta A$, and $\langle\cdot\rangle$ be an average over both quantities.
From the properties of the beta-binomial distribution we know that, for a given initial condition, we have
\begin{equation}
\left\langle \Delta A \right\rangle_{\Delta A} = \frac{\Delta N A_0}{A_0 + B_0}\,,
\label{eq:bb-mean}
\end{equation}
\begin{equation}
\textnormal{Var}[\Delta A] = \frac{\Delta N A_0 B_0 (A_0 + B_0 + \Delta N)}{(A_0 + B_0)^2 (A_0 + B_0 + 1)}\,.
\label{eq:bb-var}
\end{equation}

For the mean of $\left\langle x\right\rangle$, one obtains
\begin{equation}
\nonumber\langle x \rangle \stackrel{\eqref{eq:bb-mean}}{=} \left\langle x_0 \right\rangle_0 = \bar x_0\,.
\label{eq:supplement-exact-mean}
\end{equation}
Hence, the average composition is exactly conserved throughout the time evolution of the populations.
In other words, the stochastic process is a martingale.

For the variance we obtain
\begin{eqnarray}
\textnormal{Var}[x] &=& \left\langle \left(\frac{A_0+\Delta A}{A_0 + B_0 + \Delta N}\right)^2\right\rangle - \left\langle x_0 \right\rangle^2_0\\
&=& \left\langle \frac{A_0^2 +2 A_0 \left\langle \Delta A \right\rangle_{\Delta A} + \textnormal{Var}[\Delta A] + \left\langle\Delta A \right\rangle^2_{\Delta A}}{(A_0 + B_0 + \Delta N)^2} \right\rangle_0 - \left\langle x_0 \right\rangle^2_0\\
&\stackrel{\eqref{eq:bb-mean} }{=}& \left\langle \left(\frac{A_0}{A_0 + B_0} \right)^2 + \frac{\textnormal{Var}[\Delta A]}{(A_0 + B_0 + \Delta N)^2}\right\rangle_0 - \left\langle x_0 \right\rangle^2_0\\
&\stackrel{\eqref{eq:bb-var}}{=}& \textnormal{Var}[x_0] + \left\langle\frac{\Delta N A_0 B_0}{(A_0 + B_0)^2 (A_0 + B_0 + \Delta N)^2 (A_0 + B_0 + 1)} \right\rangle_0\\
&=& \textnormal{Var}[x_0] + \left\langle x_0 (1 - x_0)\right\rangle_0 \left\langle \frac{1}{N_0 + 1} \frac{\Delta N}{N_0 + \Delta N}\right\rangle_0\,.\label{eq:supplement-variance-part}
\end{eqnarray}
For long times (i.e., $\Delta N\gg1$), $\Delta N + N_0 \simeq \Delta N$ and \eqref{eq:supplement-variance-part} reduces to
\begin{equation}
\textnormal{Var}[x] \rightarrow \textnormal{Var}[x_0] + \left\langle \frac{1}{N_0+1} \right\rangle_0 \langle  x_0(1-x_0)\rangle_0\,.
\label{eq:supplement-exact-variance}
\end{equation}
The argument up to here is completely independent of the particular choice of initial conditions.
If the initial distribution is known, we may even make the value of the variance more explicit.
In particular, consider the distribution we obtain from experiments: in each well, $N_0$ is Poisson-distributed with mean $\bar N_0$.
Then one gets
\begin{equation}
\left\langle \frac{1}{N_0+1}\right\rangle_0 = \frac{1 - \e^{-\bar N_0}}{\bar N_0}\,.
\end{equation}
Within each well of (random) size $N_0$ there is an initial random number $A_0$ of $A$-individuals, which follows a Binomial distribution with parameters $N_0$ and $\bar x_0$.
For this choice of distribution, it is possible that $N_0=0$, which would lead to an undetermined value of $x_0=A_0/N_0$, and therefore also for the average $\left\langle x_0\right\rangle$.
We can solve this problem by redefining $x_0$:
\begin{equation}
x_0:=\begin{cases}
\bar x_0 & ,\, N_{0}=0\\
\frac{A_{0}}{N_{0}} & ,\, \textnormal{otherwise}
\end{cases}
\end{equation}
so that $x_0$ and its average have definite values, and $\langle x_0 \rangle_0 = \bar x_0$.
With this we can compute the second moment of $x_0$:
\begin{equation}
\langle x_0^2\rangle_0 = \sum_{N_0 = 1}^\infty \e^{-\bar N_0} \frac{\bar N_0^{N_0}}{N_0!}\left\lbrace\sum_{A_0 = 0}^{N_0} \binom{N_0}{A_0} \bar x_0^{A_0} (1-\bar x_0)^{N_0 - A_0} \frac{A_0^2}{N_0^2}\right\rbrace + \bar x_0^2 \e^{-\bar N_0}\,.
\end{equation}
The sum inside the braces can be solved using exponential and binomial series and yields
\begin{equation}
\langle x_0^2\rangle_0 = \bar x_0^2 + \bar x_0 (1 - \bar x_0) \e^{-\bar N_0} \sum_{N_0 = 1}^\infty \frac{\bar N_0^{N_0}}{N_0! N_0}\,.
\end{equation}
The remaining series is an exponential integral ($\mathrm{Ei}$), and therefore
\begin{equation}
\textnormal{Var}[x_0] = \bar x_0 (1 - \bar x_0) \e^{-\bar N_0} \left[\mathrm{Ei}(\bar N_0) - \gamma - \ln (\bar N_0) \right] =: \bar x_0 (1 - \bar x_0) \varphi(\bar N_0) \,,
\end{equation}
where we defined $\varphi(\bar N_0) := \e^{-\bar N_0} \left[\mathrm{Ei}(\bar N_0) - \gamma - \ln (\bar N_0) \right]$.
Then the variance of $x$ reads
\begin{eqnarray}
\textnormal{Var}[x] &=& \textnormal{Var}[x_0] + \frac{1 - \e^{-\bar N_0}}{\bar N_0} \langle x_0 (1-x_0)\rangle\\
&=& \bar x_0 (1 - \bar x_0) \left[\varphi(\bar N_0) + \frac{1 - \e^{-\bar N_0}}{\bar N_0} \left(1 - \varphi(\bar N_0)\right)\right]\,.
\end{eqnarray}
For large $\bar N_0$, through an asymptotic expansion \cite{van_zelm_wadsworth_improved_1965},
\begin{equation}
\mathrm{Ei} \simeq \frac{1}{\bar N_0} \mathrm{e}^{\bar N_0} \sum_{m=0}^{\bar N_0 - 1} m!\bar N_0^{-m} - \frac{1}{3} \sqrt{\frac{2\pi}{\bar N_0}}\,,
\end{equation}
$\varphi(\bar N_0)$ can be approximated by
\begin{equation}
\varphi (\bar N_0) \simeq \frac{1}{\bar N_0}\sum_{m=0}^{\bar N_0 - 1} m!\bar N_0^{-m} - \mathrm{e}^{-\bar N_0}\left[\frac{1}{3}\sqrt{\frac{2\pi}{\bar N_0}} - \gamma - \ln (\bar N_0) \right]\,.
\end{equation}

\noindent To leading order in $\bar N_0$, then, the variance of $x$ becomes

\begin{equation*}
\textnormal{Var}[x] = \bar x_0 (1 - \bar x_0)\frac{2}{\bar N_0}\,,
\end{equation*}

\noindent in perfect agreement with our approximate results based on Master equations (Eq. (2) in main text, see also below).

\section*{Approximate calculations for the time evolution of the distribution moments}

Using the Master Equation for the number of individuals of each strain (1), we are able to obtain the time evolution of the first three moments of the distribution of $x$.
Equation (1) is sometimes called ``Simple Growth Equation'' and can be exactly solved (see, for example,\cite{allen_introduction_2003})
using generating functions like

\begin{align}
  F(a,t) := \sum_{N_A} a^{N_A}  \: P(N_A,t).
\end{align}

To approximate the time evolution of the first three moments of $x$, however, we do not need the full solution, but only the first three moments of $N_A$ and $N_B$.
To this end, we insert the Master Equation (Eq. (1) in main text) in the definition of the generating function to get the time derivative for $F(a,t)$:

\begin{align}
\label{eq:dtF}
 \frac{d}{dt} F(a,t) = \left(-a+a^2\right) \partial_a F(a,t).
\end{align}

\noindent To obtain the time evolution of the $n$th moment, we apply the $n$th derivative with respect to $a$ on both sides of equation~\eqref{eq:dtF}, and solve for the corresponding moment.
For the first three moments, the solution is

\begin{align}
\label{eq:mom1}
 \langle N_A \rangle &= \mathrm{e}^{t} K_1\,, \\
 \label{eq:mom2}
 \langle N_A^2 \rangle &= \mathrm{e}^{ t} (\mathrm{e}^{ t} -1)K_1 + \textnormal{e}^{2 t} K_2\,, \\
 \label{eq:mom3}
 \langle N_A^3 \rangle &= \mathrm{e}^{ t} \left(-3 \mathrm{e}^{ t}+2 \mathrm{e}^{2t}+1 \right)K_1  + 3 \mathrm{e}^{2 t} \left(\mathrm{e}^{ t}-1\right) K_2 + \mathrm{e}^{3  t} K_3 \,.
\end{align}

\noindent $K_1, K_2, K_3$ are integration constants, which depend on the initial conditions.
We consider the case of Poisson initial conditions.
This means that the initial number of $A$ is Poisson-distributed with mean value $\bar N_{A,0}$,

\begin{align}
  \langle N_A(t=0) \rangle \overset{!}{=} \bar N_{A,0}\,,
\end{align}

\noindent and, since for the Poisson distribution the variance equals the mean, we get

\begin{align}
 \textnormal{Var} \, N_A(t=0)  
 \overset{!}{=} \bar N_{A,0}\,.
\end{align}
Employing these conditions in the solutions of the differential equations we found in Eq. \eqref{eq:mom1} and \eqref{eq:mom2}, we get

\begin{eqnarray}
\label{eq:momA12}
\langle N_A \rangle &=& \mathrm{e}^{t}\bar N_{A,0}\,,\\
\textnormal{Var}\,N_A &=& \mathrm{e}^{t}(2\mathrm{e}^{t}-1)\bar N_{A,0}.
\end{eqnarray}
By the known properties of the Poisson distribution, the skewness of our initial distribution equals to $1/\sqrt{\bar N_{A,0}}$.
Using Eqs. \eqref{eq:mom3}, \eqref{eq:momA12}, and the definition of the skewness, we obtain the general time evolution of the skewness

\begin{align}
\label{eq:momA3}
 v(N_A) &= \frac{\bar N_{A,0} \left(6 e^{2 t} - 6 e^{ t} + 1\right) e^{ t}}{\left(\bar N_{A,0} \left(2 e^{ t} - 1\right) e^{ t}\right)^{3/2}}\,.
\end{align}
For $N_B$, the calculations are analogous.
Note also that all calculations were exact so far.

With the moments of $N_A$ and $N_B$ we can find the (approximate) time evolution of variance and skewness of $x=N_A/(N_A+N_B)$.
For the mean of $x$ we have already seen in the exact calculation (see Eq.~\eqref{eq:supplement-exact-mean}) that it does not change with time, and hence its time evolution is already known.

To calculate the time evolution of the variance of $x$, we consider $x$ as a function of $N_A$ and $N_B$:

\begin{equation}
 x(N_A,N_B) = \frac{N_A}{N_A + N_B}.
\end{equation}
Using the time independence of the mean ($\langle x(N_A,N_B) \rangle = x(\langle N_A\rangle,\langle N_B\rangle)$), a bivariate Taylor expansion around $(\langle N_A\rangle,\langle N_B\rangle)$, and the time evolution of the moments, Eqs. \eqref{eq:momA12} and \eqref{eq:momA3}, we get for the variance of $x$:

\begin{align}
\textnormal{Var}\,x 
 &= \langle [x(N_A,N_B) - \langle x(N_A,N_B) \rangle]^2 \rangle \\
 &= \left\langle \left[ x(N_A,N_B) -  x(\langle N_A\rangle,\langle N_B\rangle) \right]^2 \right\rangle \\
 &= \langle [ x_{N_A}'(\langle N_A \rangle, \langle N_B \rangle) (N_A - \langle N_A \rangle) +\\
 &\quad + x_{N_B}'(\langle N_A \rangle, \langle N_B \rangle) (N_B - \langle N_B \rangle) + \mathcal{O} \left( N_A^{-2},N_B^{-2} \right) ]^2 \rangle \\
 &=  \frac{\langle N_B \rangle ^2}{\langle N\rangle^4} \textnormal{Var}\,N_A + \frac{\langle N_A \rangle^2}{\langle N\rangle^4} \textnormal{Var}\,N_B + \mathcal{O} \left( N_A^{-2},N_B^{-2} \right)\\
 &= \frac{(2 - \mathrm{e}^{- t})}{N_0^4}  N_{B,0} N_{A,0}\left( N_{A,0} +  N_{B,0} \right) + \mathcal{O} \left( N_A^{-2},N_B^{-2} \right)\\
 &= \frac{2-\mathrm{e}^{- t}}{\bar N_0} \bar x_0 (1-\bar x_0) \\
 \label{eq:supp-Varx}
 &\underset{t \rightarrow \infty}{\longrightarrow} \frac{2}{\bar N_0} \bar x_0 (1-\bar x_0)
\end{align}
From this we obtained Eq. (2) in main text. For infinite times the approximate result for the variance matches the exact one of Eq.~\eqref{eq:supplement-exact-variance}.

The skewness of the $x$ distribution is calculated analogously:

\begin{align}
 v(x) 
 &= \left\langle \left( \frac{x(N_A,N_B) -  x(\langle N_A \rangle,\langle N_B\rangle) }{\sqrt{\textnormal{Var}\,x}} \right)^3 \right\rangle \\
 &= \frac{x_{0} e^{- 2  t} \left(12 x_{0}^{2} e^{2  t} - 12 x_{0}^{2} e^{ t} + 2 x_{0}^{2} \right)}{N_{0}^{2} \left(\frac{x_{0}}{N_{0}} \left(- 2 x_{0} e^{ t} + x_{0} + 2 e^{ t} - 1\right) e^{-  t}\right)^{1.5}} \nonumber \\ 
 &\; \; +\frac{x_{0} e^{- 2  t} \left(- 18 x_{0} e^{2  t} + 18 x_{0} e^{ t} - 3 x_{0} + 6 e^{2  t} - 6 e^{ t} + 1\right)}{N_{0}^{2} \left(\frac{x_{0}}{N_{0}} \left(- 2 x_{0} e^{ t} + x_{0} + 2 e^{ t} - 1\right) e^{-  t}\right)^{1.5}} + \mathcal{O} \left( N_A^{-2},N_B^{-2} \right)\,. 
 \label{eq:supp-skew}
\end{align}

\section*{Comparison of initial and steady-state distributions of $\mathbf x$, and entropy of the steady state distribution conditioned on the initial one.}

\begin{figure}[!htbp]
\center\includegraphics[width=0.7\columnwidth]{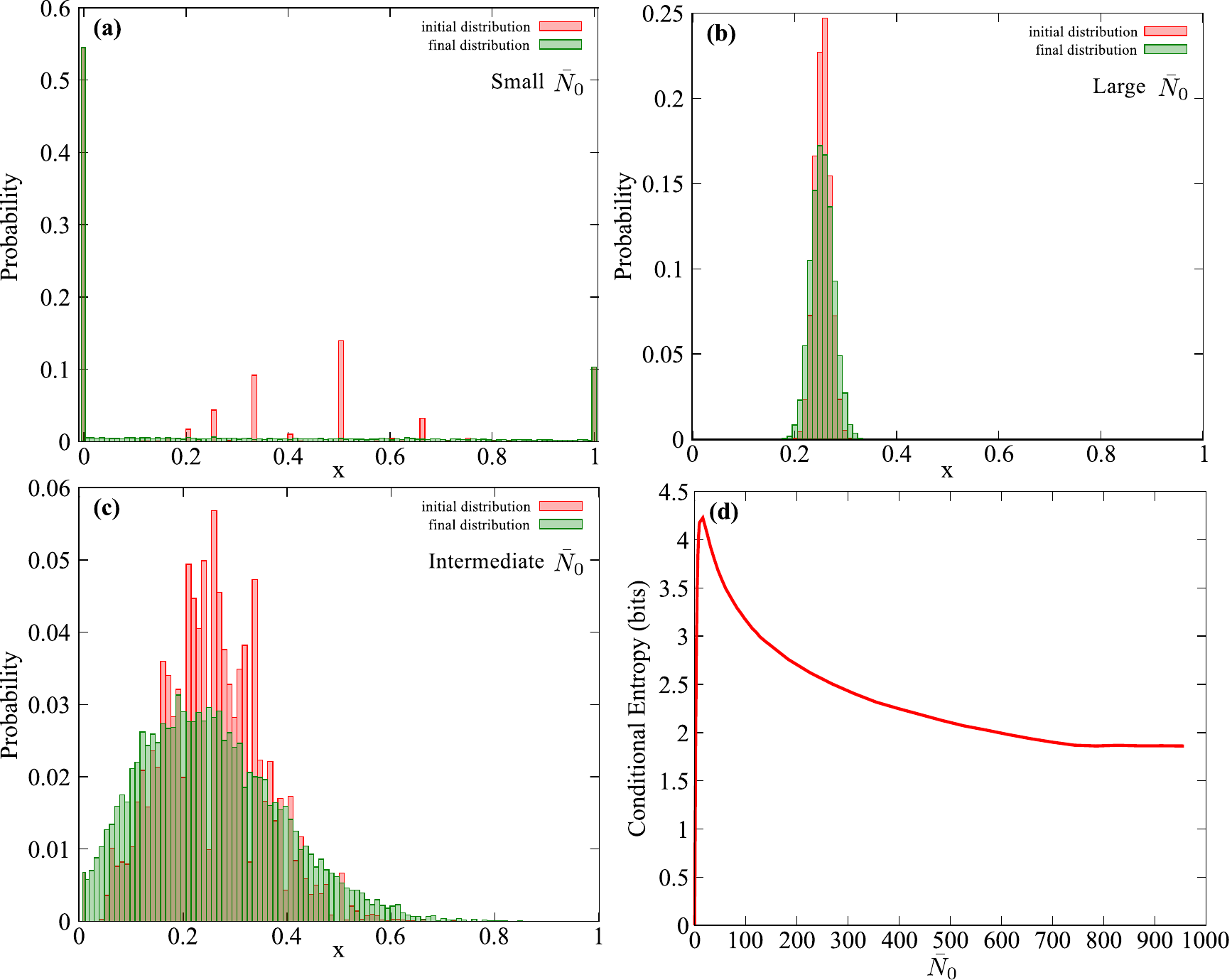}
\caption{{\bf Initial and steady state distributions, relative entropy.} \textbf{Panels (a),(b),(c)}: Initial and final distributions of $x$ for three regimes of $\bar N_0$. When $\bar N_0$ is very small or very large (panels \textbf{(a)} and \textbf{(b)}), the evolutionary fate of the population is largely determined by the initial population sampling. Therefore, the initial distribution (red bars) and the steady-state one (green bars) look qualitatively very similar.
For intermediate values of $\bar N_0$, however, population growth becomes more important, and the distributions look very different.
The amount of composition values the population can access through growth can be quantified looking at the ``unpredictability'' of the steady-state composition, once the initial one is known: the more unpredictable, the more are made accessible by growth.
Mathematically, the measure for this is called \emph{conditional entropy}: the higher the entropy, the more unpredictable the outcome.
Panel \textbf{(d)} shows the conditional entropy as function of $\bar N_0$.
Indeed, very small or very large initial populations experience little to no additional noise from growth, while in populations with intermediate values of $\bar N_0$ ($\bar N_0 \simeq 15$) growth is a major source of demographic noise.
(Parameter values: $\bar N_0=2$ \textbf{(a)}, $\bar N_0=2000$ \textbf{(b)}, $\bar N_0=20$ \textbf{(c)}; $\bar x_0 = 0.25$ in all panels)\label{fig:entropy}
}
\end{figure}

We simulate an ensemble of populations starting from Poisson initial conditions, and track their time evolution until the $x$ distribution freezes.
Once it freezes, we can build a joint histogram of initial and final compositions, which approximates the joint distribution $P_{\textnormal{joint}}(x_0,x_f)$.
From $P_{\textnormal{joint}}$ we can obtain the initial and final distributions as its marginal distributions, integrating over all values of $x_f$ and $x_0$, respectively.
The joint information (Shannon) entropy is defined as \cite{shannon_mathematical_1963}

\begin{equation}
H_{\textnormal{joint}}(x_0,x_f) = - \int_0^1 \mathrm{d}x_0\mathrm{d}x_f P_{\textnormal{joint}}(x_0,x_f) \log(P_{\textnormal{joint}}(x_0,x_f))\,.
\end{equation}
The marginal entropies $H(x_0)$ and $H(x_f)$ are defined, analogously, through integrals  only of $P(x_0)$ over $x_0$, and $P(x_f)$ over $x_f$, respectively.
The conditional entropy of the final distribution given the initial is defined as
\begin{equation}
H(x_f | x_0) = H_{\textnormal{joint}}(x_0,x_f) - H(x_0)\,.
\end{equation}
It measures the amount of information necessary to describe the final distribution, once all information about the distribution of $x_0$ is known.
Therefore, $H(x_f | x_0)$ provides a measure of how entropic (or ``noisy'') growth itself is \cite{bialek_biophysics:_2012}---or, in other words, how many different final compositions are possible given the initial condition.
Figure \ref{fig:entropy}\textbf{(d)} shows $H(x_f | x_0)$ from repeated simulations, all with the same initial distribution form, the same $\bar x_0$, but different $\bar N_0$.
For very small $\bar N_0$ (of the order of one or two individuals) the group formation almost completely determines the fate of populations: most populations start fixated, many with just a single founder individual, and the composition of each well remains the same during growth.
The path followed by $x$ in each population during time is a straight line, as the compositions stay constant.
Therefore, $x$ for different populations follow in time paths that do not cross or ``mix''.
Growth produces very little demographic noise, and its conditional entropy tends to zero.
For very large $\bar N_0$ (of the order of a few hundreds), the group sampling is again central to 
determine the final distribution.
Very large populations, in fact, all start with similar compositions (according to the Law of Large Numbers), and their compositions are difficult to change, as each individual event has little impact.
The composition distribution changes very little before freezing; time evolution paths of different populations ``mix'' very little.
Entropy in this regime saturates for increasing initial sizes, and is rather low.
Between the small size regime (where paths do not ``mix'') and the large size regime (where size limits ``mixing''), we find a window where populations are small enough to significantly change their composition, but also large enough to not start fixated.
This is the region where the conditional entropy peaks, and growth is the most important in determining the final distribution.

Intuitively, the difference in variance between initial and final distribution could provide an alternative measure of the noise introduced by growth.
However, of all $x$ distributions between $0$ and $1$ with fixed $\bar x_0$, the one with maximal variance is the one for which $x$ is only $0$ or $1$, i.e., when all populations start off fixated.
In this case, the compositions never change during growth and the variance stays constant.
Moreover, independently on the choice of initial distribution, the difference between initial and steady-state variance decreases for increasing $\bar N_0$ (see Eq.\eqref{eq:supplement-exact-variance}).
Therefore, all considerations on noise sources based on variance would indicate that growth matters more when initial populations are smaller, in contrast with our observations.

 \begin{table}[!htbp]
 	\center
 	\begin{tabular}{|l|c|c|}
 		\hline
 		& \textit{Drosophila} & \textit{P.Putida}\\\hline
 		
 		\textbf{\# of populations} & 107 & 120\\\hline
 		
 		\textbf{Initial pop. size} & 16 & $\sim$ 10\\\hline
 		
 		\textbf{Max. \# of generations} & 19 & 16\\\hline
 		
 		\textbf{Pop. size} & Constant & Growing\\\hline
 		
 		\textbf{Outcome} & Increasing number of fixations & No fixation, freezing\\
 		\hline
 	\end{tabular}
 	\caption{{\bf Comparison between results from our experiments and those in [9].}While experiments for constant-sized populations of \textit{Drosophila} observe significant fixations within the first tens of generations, we instead observe freezing of the probability distribution for the population composition, without any fixation.
 	}
 \end{table}

\begin{figure}
\center\includegraphics[width=0.6\columnwidth]{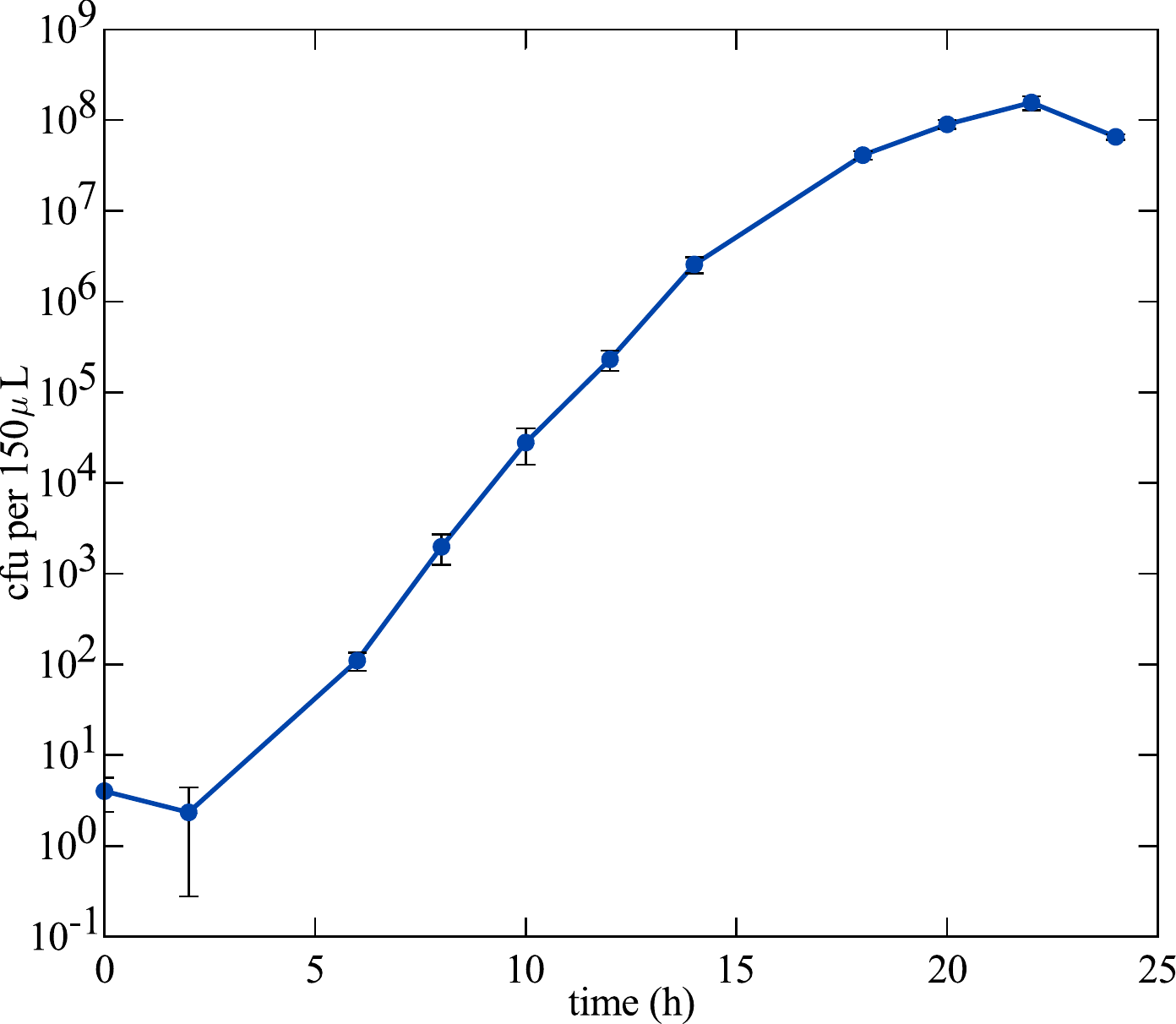}
\caption{{\bf Growth curve of a mixed population.}
The population consists of pyoverdine producer (\textit{P. putida} KT2440) and non-producer (\textit{P. putida} 3E2) under non-selective (iron replete) conditions.
Individual precultures of the strains were mixed and diluted in iron replete medium to yield $\bar N_0=4$ (in 150 $\mu$L), and $\bar x_0=0.5$.
Cells were grown aerobically at 30{\textdegree}C for 24 hours.
The dots represent the mean $N(t)$ of three independent replications, the bars the corresponding standard deviation.
After a lag phase of about 2 hours, the cells start to grow exponentially and reach the stationary phase after about 14 h of growth.
For the non-selective growth experiments used to test the predictions of the P\'olya urn model, cells were grown for 11.5 h to ensure exponential growth conditions.
}
\end{figure}

 \begin{table}[!htbp]
 	\center
 	\begin{tabular}{|r|c|c|c|c|}
 		\hline
 		& \multicolumn{2}{|c|}{\textbf{Specific growth rate (h$^{-1}$)}} & \multicolumn{2}{|c|}{\textbf{Fluorescence per cell (a.u.)}}\\
 		\hline
 		\textbf{Iron conc. ($\mathbf{\mu}$M)} & \textit{KT2440}  & \textit{3E2} & \textit{KT2440}  & \textit{3E2}\\
 		\hline
 		0                           & 0.058$\pm$0.006 & \textit{no growth} & 244.00 $\pm$ 21.3   & 0 $\pm$ 0\\
 		\hline
 		200                         & 0.152$\pm$0.026 & 0.146$\pm$0.017 & 1.56 $\pm$ 0.27 & 0.93 $\pm$ 0.10\\\hline
 	\end{tabular}
 	\caption{{\bf Comparison of growth and pyoverdine production per cell of \textit{P. putida} KT2440 and 3E2.} Separate cultures of producer (\textit{P. putida} KT2440) and non-producer (\textit{P. putida} 3E2) were grown in iron-limiting (no addition of FeCl$_3$) and iron-replete medium (addition of 200 $\mu$M FeCl$_3$) at 30\textdegree C. The cell density was measured at 600 nm, and specific growth rates were calculated from density values of the exponential phase. The pyoverdine production was determined by fluorescence emission measurements (excitation 400 nm, emission at 460 nm). The pyoverdine production per cell represents the ratio of pyoverdine fluorescence and optical density measured after 24 h of growth.
 	The values in the table are averages over a minimum of five experiments, with the corresponding standard deviation.
 	The fluorescence value for the non-producing mutant in iron-limiting medium is 0 because the culture failed to grow.}
 \end{table}

\begin{figure}
\center\includegraphics{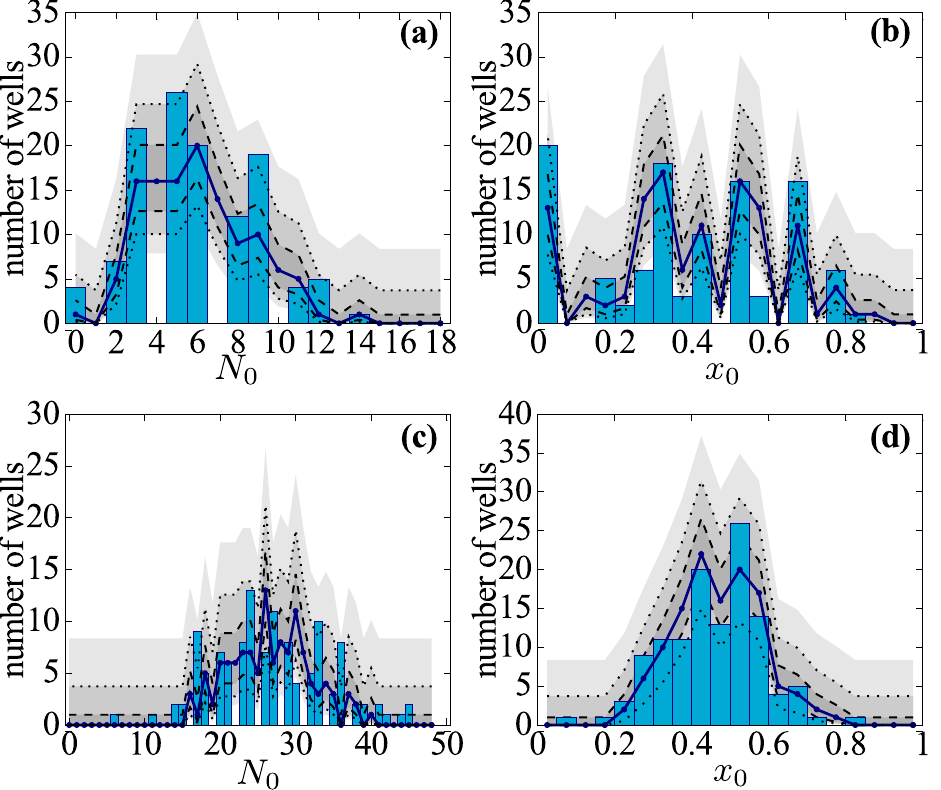}
\caption{{\bf Additional initial conditions measurements.}
The experimental distributions (bars) are measured from 120-well ensembles, the average $N_0$ and $x_0$ from those sets the parameters for the simulated distributions.
The theoretical average distribution (solid line) is the average of the same distributions generated for 84 sets of 120 wells.
Using that average we calculate three Wilson binomial confidence intervals (gray areas).
Experiments and theory agree within statistical error: the distribution of sizes (panels \textbf{(a)} and \textbf{(c)}) follows a Poisson distribution.
The raggedness of the distribution of $x$ for at small $\bar N_0$ (see panel \textbf{(b)} and Fig. 3\textbf{(b)} in main text) is due to a small size effect: since $x$ must be a simple fraction, when $N_0$ is small only a few values are available (see main text). This effect disappears for average initial sizes $\bar N_0 \simeq 10$ (see panel \textbf{(d)}).
Parameter values: $\bar N_0=5.75$, $\bar x_0=0.43$ \textbf{(a)} and \textbf{(b)}; $\bar N_0=26.49$, $\bar x_0=0.45$ \textbf{(c)} and \textbf{(d)}.}
\end{figure}
\end{document}